\documentclass[twocolumn,floatfix,showpacs]{revtex4}
\usepackage{amsfonts}
\usepackage{amsmath}
\usepackage{amssymb}
\usepackage{mathrsfs}
\usepackage[english,activeacute]{babel}
\usepackage{graphicx}
\usepackage{subfigure}
\usepackage{multirow}
\usepackage{epsfig}
\usepackage{float}

 \usepackage{color}

\begin{document}

\title{{\small Zeitschrift f. Naturforschung A 73(9), pp. 805-814 (2018)%\\ arXiv: 1604.04807v3
}\\ \bigskip
{\Large Two integrable classes of Emden-Fowler equations\\ with applications in astrophysics and cosmology}}

\author{Stefan C.  Mancas}\email{mancass@erau.edu}
\affiliation{Department of Mathematics, Embry-Riddle Aeronautical University, Daytona Beach, FL 32114-3900, USA}
%\ead{mancass@erau.edu}

\author{Haret C. Rosu}\email{hcr@ipicyt.edu.mx (Corresponding-author)}
%\author{xxx}\email{yyy}
\affiliation{IPICyT, Instituto Potosino de Investigacion Cientifica y Tecnologica,\\
Camino a la presa San Jos\'e 2055, Col. Lomas 4a Secci\'on, 78216 San Luis Potos\'{\i}, S.L.P., Mexico}%

\begin{abstract}
\noindent  We show that some Emden-Fowler (EF) equations encountered in astrophysics and cosmology belong to two EF integrable classes of the type  $d^2z/d{\chi}^2=A \chi ^{-\lambda-2}z^n$  for $\lambda=(n-1)/2$ (class one), and $\lambda=n+1$ (class two). We find their corresponding  invariants which reduce them to first order nonlinear ordinary differential equations. Using particular solutions of such EF equations, the two classes are set in the autonomous nonlinear oscillator form $d^2\nu/dt^2 +ad \nu/dt +b(\nu-\nu^n)=0$, where the coefficients $a,b$ depend only on $\lambda, n$. For both classes, we write closed-form  solutions in parametric form. The illustrative examples from astrophysics and general relativity correspond to two $n=2$ cases from class one and two, and one $n=5$ case from class one, all of them yielding Weierstrass elliptic solutions. It is also noticed that when  $n=2$, EF equations can be studied using the Painlev\'e reduction method, since they are a particular case of equations of the type $d^2z/d{\chi}^2=F(\chi)z^2$, where $F(\chi)$  is the Kustaanheimo-Qvist function.
\\

\pacs{02.30.Hq, 04.20.Jb, 02.30.Ik \hfill {\bf arXiv: 1604.04807v3}}
%ode's; exact solutions, integrable systems


\end{abstract}

\maketitle

\section{Introduction}

Several particular cases of the homogeneous ordinary differential
equations (ODEs) of the form
\begin{equation}\label{ef0}
\frac{d^2y}{dx^2}+\frac{N}{x}\frac{dy}{dx}+f(x,y)=0~,
%=A x^{-\lambda-2}y^n~.
\end{equation}
where $N>0$, $x\geq 0$, and $f(x,y)$, a nonlinear function, have proved to be extremely significant in fundamental topics of astrophysics,
atomic physics, and other areas. Their first occurrence was in astrophysics in the particular
case $N=2$ and $f(x,y)$ a monomial power of $y$
\begin{equation}\label{ef01}
\frac{d^2y}{dx^2}+\frac{2}{x}\frac{dy}{dx}+y^n=0~,
%=A x^{-\lambda-2}y^n~.
\end{equation}
or, in self-adjoint form
\begin{equation}\label{ef02}
\frac{d}{dx}\left(x^2\frac{dy}{dx}\right)+x^2y^n=0~,
%=A x^{-\lambda-2}y^n~.
\end{equation}
with $n=3$ the most natural case, which is known as the Lane-Emden equation.
For standard initial conditions $y(0)=1$ and $y'(0)=0$, there are solutions of~(\ref{ef01}) representing the Newton-Poisson gravitational potential of stars, such as the Sun, considered as spheres filled with a polytropic gas. This has been first shown in the famous book of Emden \cite{Emden} written at the beginning of the 20th century. Emden's book not only summarized the first four decades of research on the self-gravitating stars initiated by Lane in 1870 \cite{Lane}, but also introduced one of the first singular Cauchy problem as represented by ~(\ref{ef01}), and moreover stimulated further remarkable studies of a whole generation of renowned astrophysicists and mathematicians, such as Eddington \cite{Edd}, Fowler \cite{Fow30}, and Milne \cite{Mil30}. Furthermore,
for negative values of the parameter $n$, there are important applications in the theory of spherical nebulae and clusters \cite{RM54}, and in the Emden-Fowler form to more terrestrial areas such as pseudoplastic fluids and large deflections of membranes in mechanics \cite{JM94}.

A `two-parameter' generalization of the Emden equation was introduced and systematically studied by Fowler \cite{Fow30}
\begin{equation}\label{ef03}
\frac{d}{dx}\left(x^2\frac{dy}{dx}\right)+x^\lambda y^n=0
%=A x^{-\lambda-2}y^n~.
\end{equation}
and is currently known as the Emden-Fowler (EF) equation, while  %because of Fowler's systematic investigation in 1930 \cite{Fow30}.
the Lane-Emden equation is the particular case $\lambda=2$. By the change of variables $y(x)=z(\chi)$, where $\chi=1/x$, the EF equation can be written in the standard non-adjoint form
\begin{equation}\label{ef03b}
\frac{d^2z}{d\chi^2}+ \chi^{-\lambda-2}z^n=0~,
\end{equation}
which is the usual form encountered in the literature.

As for the `three-parameter' generalization
\begin{equation}\label{ef04}
\frac{d}{dx}\left(x^\rho\frac{dy}{dx}\right)\pm x^\lambda y^n=0
%=A x^{-\lambda-2}y^n~,
\end{equation}
changes of both dependent and independent variables, thumbnail sketched by Bellman \cite{B53} from Fowler's work, reduce it to the standard form
\begin{equation}\label{ef04b}
\frac{d^2z}{d\chi^2}\pm \chi^{\sigma}z^n=0~,
\end{equation}
which is equivalent to~(\ref{ef03b}) and so nothing essentially new is added.
The Thomas-Fermi model \cite{Thomas,Fermi} for the electrostatic field in the bulk of a heavy atom originates from Coulomb-Poisson considerations and is expressed by a self-adjoint form of the kind given in~(\ref{ef04}) although for a non-integer $n$. In the literature, one can also find an interesting paper on the exact solutions of a four-parameter generalization of the Lane-Emden equation, although one of the parameters can be scaled to $\pm 1$ by scale transformations \cite{gh}.

\medskip

In this work, we are concerned with the EF equation in the standard form
% slightly modified by a constant factor $A$ in the non operatorial part
\begin{equation}\label{ef03c}
\frac{d^2z}{d\chi^2}=A\chi^{-\lambda-2}z^n~,
\end{equation}
for which we provide a detailed discussion of two important integrable cases corresponding
to $\lambda=\frac{n-1}{2}$ and $\lambda=n+1$. These choices reduce the EF equation to single
parameter forms which, however, are the most encountered in astrophysical applications. In some sense,
the solution method that we here provide for these cases, in particular for $n=5$ and for $n=2$ for
which we also add the Painlev\'e reduction, may be considered as complementary to the method
of power series solutions which is widely used in the theoretical astrophysics of the general relativistic
isotropic fluid stars, see \cite{hm} and references therein. In the past, several authors have dealt with the integration of the Emden-Fowler and the generalized Emden-Fowler equations through the invariant variational principles and group-invariant techniques \cite{Bhut}, or through admissible functional transformations to  Abel's equation of the second kind \cite{Pana1,Pana2}. Moroever, the authors in \cite{Car} developed a theory of quasi-Lie schemes to  investigate several equations of Emden type  where they obtain constants of motion by means of particular solutions, which requires a priori knowledge of some particular solution that enables transformation from an Emden equation into a Lie system \cite{Blu,Pete}. In addition, in a paper by Leach and collaborators \cite{LMM92} on generalized EF equations of the form $y''+f(x)y^n=0$, closed-form solutions have been obtained for some of the special cases of $f(x)$ for which the equation possesses one or two Lie point symmetries.

Before proceeding with the main part of the paper, we mention the two particular cases of~(\ref{ef03c}).
First, for $\lambda=0$ we obtain
\begin{equation}\label{eq11a}
\frac{d^2z}{d\chi^2}=A\chi^{-2}z^n~,
\end{equation}
which is transformed to
\begin{equation}\label{eq11b}
\frac{d^2z}{dt^2}-\frac{dz}{dt}=Az^n
\end{equation}
by using $t =\ln \chi$.
% which was also discussed in detail in 1914 by Fowler \cite{Fow14}.
The second particular case is $\lambda=n-1$ which leads to
 \begin{equation}\label{eq111a}
 \frac{d^2z}{d\chi^2}=A\chi^{-n-1}z^n~,
 \end{equation}
and using again  $t =\ln \chi$  simplifies to
 \begin{equation}\label{eq1002}
\frac{d^2z}{dt^2}-\frac{dz}{dt}=A\chi ^{-n+1}{z}^n~.
 \end{equation}
Moreover, by letting $z=\theta e^t$, we obtain
 \begin{equation}\label{eq112a}
\frac{d^2\theta}{dt^2}+\frac{d\theta}{dt}=A{\theta}^n~.
 \end{equation}
 Equations (\ref{eq11b}) and (\ref{eq112a}) were discussed previously in detail by Fowler \cite{Fow30,Fow14} and will not be of interest here.

\section{Self-adjoint and invariant forms}

 The self-adjoint form of~(\ref{ef03c}) is obtained by changing the variables %using Kamke's substitutions \cite{Kamke} %using the  canonical variables
 according to   $z(\chi)=\eta(\xi)$ and $\xi=\frac{1}{\chi}$,
 which leads to
 \begin{equation}\label{EFnormal}
\frac{1}{\xi^2}\frac {d}{d \xi}\left(\xi^2 \frac{d\eta}{d\xi}\right)=A\xi^{\lambda-2}\eta^{n}~.
 \end{equation}

On the other hand, an invariant form of~(\ref{ef03c}) can be obtained using the transformation $z(\chi)=\frac{w(s)}{s}$, where $s=\frac 1 \chi$, which leads to an EF equation with a different power of the independent variable containing both parameters
%..........................
\begin{equation}\label{eq1}
\frac{d^2 w}{ds^2}=A s^{\lambda -1-n}w^n~.
\end{equation}

One can then infer that, if $\lambda=\frac{n-1}{2}$,~(\ref{eq1}) is invariant under the transformation $z  \leftrightarrow \frac{w(s)}{s}~,~ s\leftrightarrow\frac 1 \chi$ since it coincides with~(\ref{ef03c}), while for $\lambda=n+1$, the power of $s$ is zero. Thus, it does not appear to be a coincidence that in 1984 when Rosenau \cite{ros} was interested in the integration of~(\ref{ef03c}), he provided two algebraic methods of integration when the same conditions between the powers of the variables  are satisfied:  $\lambda=\frac{n-1}{2}$ or $\lambda=n+1$. Thus,  both cases will be considered in detail, for which we will construct integrals of motions and write  parametric solutions using a general formalism.

To proceed with the general  solutions for these two classes, first let us assume that, if we have  a particular solution $z_p(\chi)$ of~(\ref{ef03c}), then  we can construct a general  solution $z(\chi)=z_p(\chi) \nu(\chi)$, where $\nu(\chi)$ must satisfy the ODE
\begin{equation}\label{eq1a1}
z_p \frac{d^2 \nu}{d \chi^2}+2 \frac{d z_p}{d \chi}\frac{d \nu}{d \chi}+\frac{d^2 z_p}{d \chi^2}(\nu-\nu^n)=0~.
\end{equation}
%For both classes the coefficients $\frac{z_{p}\,\rq{}}{z_p}$ and $\frac{z_p\,\rq{}\rq{}}{z_p}$   turn to be constants.
%%%%%%%%%%%%%%%%%%%%%%%%%%%%%%%%%%%%%%%%%
\subsection{Case $\lambda=\frac{n-1}{2}$}
Let $\lambda=\frac{n-1}{2}$ in~(\ref{ef03c}) we obtain
 \begin{equation}\label{ef2}
\frac{d^2 z}{d \chi^2}=A \chi^{-\frac{n+3}{2}}z^n~,
\end{equation}
which has a particular solution
 \begin{equation}\label{ef2a}
 z_p(\chi)=\alpha \sqrt \chi \, ,\hspace{1cm} \alpha^{n-1}=-\frac{1}{4A}~.
 \end{equation}
  To construct the  integrals of motion, we first put~(\ref{ef2}) in  self-adjoint form to obtain
 \begin{equation}\label{ef3}
\frac {d}{d \xi}\left(\xi^2 \frac{d\eta}{d\xi}\right)=A\xi^{\frac{n-1}{2}}\eta^{n}~.
\end{equation}
In 1970s, Djukic \cite{Djuk} was first to find integrals of motion for~(\ref{ef2}) using a generalized Noether theorem. He obtained
 \begin{equation}\label{99}
\xi^2 \frac{d\eta}{d\xi}\left(\xi \frac{d\eta}{d\xi}+\eta\right)-\frac{2A}{n+1}\xi^{\frac{n+1}{2}}\eta^{n+1}={\mathcal C_1}~,
 \end{equation}
 which in terms  of the original variables of~(\ref{ef2}) turns into
  \begin{equation}\label{100}
\frac{dz}{d\chi}\left(\chi \frac{dz}{d\chi}-z\right)-\frac{2A}{n+1}\chi^{-\frac{n+1}{2}}z^{n+1}={\mathcal C_1}~.
 \end{equation}
The invariant method may have advantage over the original equation because one may obtain more easily the solution by solving a first-order ODE instead of solving a second-order one.

Using the invariant form (\ref{eq1}) with $\lambda=\frac{n-1}{2}$, we obtain
  \begin{equation}\label{ef4}
\frac{d^2w}{ds^2}=A s^{-\frac{n+3}{2}}w^n~,
\end{equation}
which is the same as (\ref{ef2}). Thus, its invariant is
  \begin{equation}\label{101}
\frac{dw}{ds}\left(s\frac{dw}{ds}-w\right)-\frac{2A}{n+1}s^{-\frac{n+1}{2}}w^{n+1}={\mathcal C_2}~.
 \end{equation}
By using the original variables, the invariant (\ref{101}) is exactly (\ref{100}) up to some arbitrary integration constant.

Since we know $z_p$ according to ~(\ref{ef2a}),~(\ref{eq1a1}) reduces to
\begin{equation}\label{eq23a}
\chi^2\frac{d^2 \nu}{d \chi^2}+\chi \frac{d\nu}{d\chi}-\frac 1 4 (\nu-\nu^n)=0~.
\end{equation}
The damping term can be eliminated using $t=\ln \chi$ and gives
\begin{equation}\label{eq23b}
\frac{d^2 \nu}{d t^2}=\frac 1 4 (\nu-\nu^n)~,
\end{equation}
with one solution found by Fowler \cite{Fow}
\begin{equation}\label{er200tris}
\nu(t)=\left[2(n+1)\frac{K_1e^{-\frac{n-1}{2}t}}{\left(1+K_1e^{-\frac{n-1}{2}t}\right)^2}\right]^{\frac{1}{n-1}}~, \qquad n\neq 1~.
\end{equation}
By combining~(\ref{er200tris}) and (\ref{ef2a}), the particular solution to~(\ref{ef2}) is
\begin{equation}\label{er300tris}
z_p(\chi)=\left[\frac{-K_1(n+1)}{2A\left(1+K_1\chi^{-\frac{n-1}{2}}\right)^2}\right]^{\frac{1}{n-1}}~, \qquad n\neq 1~.
\end{equation}
%The case $n=1 \Rightarrow \lambda=0$ gives Eq. (\ref{eq11b}).
 %%%%%%%%%%%%%%%%%%%%%%%%%%%%%
\subsection{Case $\lambda=n+1$}

Letting  $\lambda=n+1$  in~(\ref{ef03c}), we get
 \begin{equation}\label{eg2}
\frac{d^2 z}{d \chi^2}=A \chi^{-(n+3)}z^n~,
\end{equation}
which has the particular solution
 \begin{equation}\label{eg2a}
 z_p(\chi)=\beta \chi^{\frac{n+1}{n-1}} \, ,\hspace{1cm} \beta^{n-1}=\frac{2(n+1)}{A(n-1)^2} ~, \qquad n\neq 1~.
 \end{equation}
To construct the integrals of motion, we will proceed in reverse order, by using first the invariant transformation, and then the canonical variables.
Using  $\lambda=n+1$ in~(\ref{eq1}), we have
\begin{equation}\label{eg3}
\frac{d^2w}{ds^2}=A w^n~.
\end{equation}

The integral of motion is easily obtained by multiplying by $w_s$ and integrating to get
\begin{equation}\label{1003}
\left(\frac{dw}{ds}\right)^2-\frac{2 A}{n+1} w^{n+1}=\mathcal{C}_3~,
\end{equation}
which in terms of the original variables of~(\ref{eg2}) becomes
\begin{equation}\label{1005}
\left(z-\chi \frac{dz}{d\chi}\right)^2-\frac{2 A}{n+1}\chi^{-(n+1)}z^{n+1}=\mathcal{C}_3~.
\end{equation}
Using the canonical variables,~(\ref{eg2}) is
\begin{equation}\label{eg4}
\frac {d}{d \xi}\left(\xi^2 \frac{d\eta}{d\xi}\right)=A\xi^{n+1}\eta^{n}~,
\end{equation}
with  invariant
\begin{equation}\label{1006}
\left(\eta+\xi \frac{d\eta}{d\xi}\right)^2-\frac{2 A}{n+1}\xi^{n+1}\eta^{n+1}=\mathcal{C}_3~.
\end{equation}

 Since we know $z_p$, according to~(\ref{eq1a1}) we obtain
\begin{equation}\label{er23}
\chi^2\frac{d^2\nu}{d\chi^2}+\frac{2(n+1)}{n-1}\chi \frac{d\nu}{d\chi}+\frac{2(n+1)}{(n-1)^2}(v-v^n)=0~,
\end{equation}
and by using $t=\ln \chi$ we obtain
\begin{equation}\label{er23a}
\frac{d^2\nu}{d t^2}+\frac{n+3}{n-1}\frac{d \nu}{dt}+\frac {2(n+1)}{(n-1)^2} (\nu-\nu^n)=0~.
\end{equation}
%This equation was also discussed in detail in the 1931 paper by Fowler \cite{Fow}.
A particular solution of this equation can be obtained using the factorization technique  of Rosu and Cornejo-P\'erez \cite{ro}.
The factored form of~(\ref{er23a}) is
\begin{equation}\label{er23bis}
\bigg[\frac{d}{dt}+\frac{n+1}{n-1}\left(1+\nu^{\frac{n-1}{2}}\right)\bigg]
\bigg[\frac{d}{dt}+\frac{2}{n-1}\left(1-\nu^{\frac{n-1}{2}}\right)\bigg]\nu=0~.
\end{equation}
This leads to
\begin{equation}\label{er23tris}
\frac{d \nu}{dt}+\frac{2}{n-1}\nu-\frac{2}{n-1}\nu^{\frac{n+1}{2}}=0
\end{equation}
which is a Bernoulli equation with the kink solution
\begin{equation}\label{er23tris1}
\nu(t)=\frac{1}{(1+K_2e^t)^{\frac{2}{n-1}}}~, \qquad n\neq 1~.
\end{equation}
By combining~(\ref{er23tris1}) and (\ref{eg2a}), the particular solution to~(\ref{eg2}) is
\begin{equation}\label{er300tridd}
z_p(\chi)=\left[\frac{2(n+1)}{A(n-1)^2}\frac{\chi^{n+1}}{(1+K_2\chi)^2}\right]^{\frac{1}{n-1}}~, \qquad n\neq 1~.
\end{equation}

%%%%%%%%%%%%%%%%%%%%%%%%%%%%%%%%%%%%%%%%%%%%%%%%%%%%%
%%%%%%%%%%%%%%%%%%%%%%%%%%%%%%%%%%%%%%%%%%%%%%%%%%%%%%
\section{Parametric solutions}
\subsection{Case $\lambda=\frac{n-1}{2}$}

Equation~(\ref{ef2}) is the same as~(4) in Section %\S
2.3.1-2 of  Polyanin\rq{}s book \cite{Pol}. Depending on $n$, there are two sets of parametric solutions, namely
\begin{eqnarray}\label{equ2}
\begin{array}{ll}
&\chi(\tau)=a B_{1}^{\,2}\exp \Theta(\tau)~,\\
&z(\tau)=b B_1 \tau \exp{\frac{1}{2}\Theta (\tau)}~,\\
&\Theta(\tau)=\int{\frac{d \tau}{\sqrt{K_3+\psi(\tau)+\frac{\tau^2}{4}}}}~,\\
& A=\left(\frac{a}{b^2}\right)^{\frac{n-1}{2}}~.
\end{array}
\end{eqnarray}
First, if $n \ne -1$, we have
$\psi(\tau)=\frac{2\tau^{n+1}}{n+1}$, while if $n=-1$, $\psi(\tau)= 2 \ln |\tau|$.

%%%%%%%%%%%%%%%%%%%%%%%%%%%%%

\subsection{Case $\lambda=n+1$}

Equation~(\ref{eg2}) is the same as~(3) in Section %\S
2.3.1-2 of  Polyanin\rq{}s book \cite{Pol}. Therefore, depending on $n$ again, there are also two sets of parametric solutions. When $n \ne -1$, we have
\begin{eqnarray}\label{equ5}
\begin{array}{ll}
&\chi(\tau)=\frac{a C_{1}^{\,n-1}}{\Theta(\tau)}~,\\
&z(\tau)=\frac{b C_{1}^{\,n+1}\tau}{\Theta(\tau)}~,\\
&\Theta(\tau)=\Theta_0+\int{\frac{d \tau}{\sqrt{1\pm \tau^ {n+1}}}}~,\\
& A=\pm \frac {n+1}{ 2}~a^{n+1}b^{1-n}~,
\end{array}
\end{eqnarray}
while for $n=-1$ we have
\begin{eqnarray}\label{equ6}
\begin{array}{ll}
&\chi(\tau)=\frac{C_1}{\Theta(\tau)}\\
&z(\tau)=\frac{b \exp{(\mp \tau^2)}}{\Theta(\tau)}\\
&\Theta(\tau)=\Theta_0+\int{\exp{(\mp \tau^2)} d \tau}\\
& A=\mp 2b^2~.
\end{array}
\end{eqnarray}

\section{Examples}
 We will present three examples, choosing $n=2$, and  $n=5$ for the  first class  while taking $n=2$ for the second class. The case $n=5$  occurs in astrophysics of stars treated as gas spheres, while for $n=2$ the Emden-Fowler equations are particular cases of equations of the type
\begin{equation}\label{eqkq2}
\frac{d^2 z}{d \chi^2}=F(\chi)z^2~,
\end{equation}
which describe perfect fluids in shear-free motion in general relativity as shown by Kustaanheimo and Qvist already in 1948 \cite{KQ}.
These authors obtained the remarkable result that~(\ref{eqkq2}) is integrable when the function $F$ takes the form
\begin{equation}\label{qvist}
F_{_{KQ}}(\chi)=(\alpha_1\chi^2+\alpha_2\chi +\alpha_3)^{-5/2}
\end{equation}
for arbitrary constants $\alpha_1$, $\alpha_2$, and $\alpha_3$, a result that was later also found by \cite{Leach} using the methods of Lie point symmetry analysis. The Emden-Fowler case from the first class corresponds to $\alpha_1=\alpha_3=0$, and $\alpha_{2}=A^{-2/5}$, while the case from the second class is obtained when $\alpha_2=\alpha_3=0$ and $\alpha_{1}=A^{-2/5}$. What Kustaanheimo and Qvist did not know was that Ince had studied in considerable detail equations of the type (\ref{eqkq2}) in his book \cite{Ince} by the method of reduction to Painlev\'e equations. For completeness, following Ince, we briefly present this reduction method here.
% %%%%%%%%%%%%%%%%%%%%%%%%%%%%%%%%%%%
\subsection{Reduction to Painlev\'e equations}
The necessary condition for  a second order equation
\begin{equation}\label{m1p}
 z_{\chi \chi}+h_2(z)z_{\chi}  +h_3(z)= 0~
\end{equation}
 for nonappearance of movable branch points is that the nonlinearities are polynomials, i.e. $h_2(z)=-(a_1z+a_0)$, and $h_3(z)=-(b_3z^3+Fz^2+b_1z+b_0)$, with all coefficients either constants or functions of the independent variable. With this assumption,
 ~\eqref{m1p} takes the form
\begin{equation}\label{ansa}
z_{\chi\chi}=(a_1z+a_0)z_\chi+b_3z^3+Fz^2+b_1z+b_0~.
\end{equation}
To proceed with the integrability of this equation, we perform the Painlev\'e transformation \cite{Pain}
\begin{equation}\label{ans}
z=\lambda(\chi)W(Z)+\mu(\chi)~,  \qquad Z=\phi(\chi)~,
\end{equation}
where the functions $\lambda$, $\nu$, $\phi$ are to be found in such a way that $W$ satisfies a Painlev\'e, or an elliptic equation. By using this ansatz in~\eqref{ansa}, we obtain
\begin{equation}\label{ansa2}
W_{ZZ}=(A W+B)W_Z+C W^3+DW^2+E W+S~,
\end{equation}
where
\begin{align}\label{triad}
%\left\{
%\begin{array}{lll}
A&=\frac{a_1\lambda}{\phi_\xi}~, \nonumber\\
B&=\frac{(a_0+a_1 \mu)\lambda \phi_\chi-2 \lambda _\chi\phi_\chi-\lambda \phi_{\chi \chi}}{\lambda (\phi _\chi) ^2}~, \nonumber \\
C&=\frac{b_3 \lambda^2}{(\phi_\chi)^2}~, \\
D&=\frac{(F+3 b_3 \mu)\lambda+a_1 \lambda_\chi}{(\phi_\chi)^2}~, \nonumber \\
E&=\frac{(b_1+2F \mu+3 b_3 \mu ^2+a_1 \mu _\chi)\lambda+(a_0+a_1 \mu)\lambda_{\chi} - \lambda_{\chi \chi}}{\lambda (\phi _\chi) ^2}~, \nonumber \\
S&=\frac{b_0+b_1 \mu+ F \mu ^2+b_3 \mu^3+(a_0+a_1 \mu)\mu_{\chi} - \mu_{\chi \chi}}{\lambda (\phi _\chi) ^2}~. \nonumber
%\end{array}
%\right.
\end{align}
Since we have the form given by \eqref{eqkq2}, $a_1=a_0=b_3=b_1=b_0=0$, and $n=2$, which gives the particular case $A=C=0$. Equation~\eqref{ansa2} reduces to
\begin{equation}
W_{ZZ}=BW_Z+DW^2+E W+S~,
\end{equation}
 which simplifies to
 \begin{widetext}
\begin{equation}\label{ansa3}
W_{ZZ}=-\frac{2 \lambda _\chi\phi_\chi+\lambda \phi_{\chi \chi}}{\lambda (\phi _\chi) ^2}W_Z+\frac{F\lambda}{(\phi_\chi)^2}W^2+\frac{2F \mu\lambda - \lambda_{\chi \chi}}{\lambda (\phi _\chi) ^2}W+\frac{ F \mu ^2 - \mu_{\chi \chi}}{\lambda (\phi _\chi) ^2}~.
\end{equation}
\end{widetext}
This will reduce further to
\begin{equation}\label{le}
W_{ZZ}=6 W^2+S
\end{equation}
by choosing
\begin{align}\label{triad2}
\left\{
\begin{array}{lll}
2\frac{\lambda_\chi}{\lambda}+\frac{\phi_{\chi\chi}}{\phi_\chi}&=0~,\\
2 \mu \lambda F &=\lambda_{\chi\chi}~,  \\
F&=6\frac{\phi_\chi^2}{\lambda}~.
\end{array}
\right.
\end{align}
$\lambda$ can be found by integrating once the first equation of the system~\eqref{triad2}
\begin{equation}\label{lam1}
\phi_{\chi}=\frac{1}{\lambda^2}~,
\end{equation}
and substituting into the last equation of the system~\eqref{triad2} to obtain
\begin{equation}\label{lam2}
\lambda(\chi)=\sqrt[5]{\frac{6}{F(\chi)}}~.
\end{equation}
Using \eqref{lam2} into \eqref{lam1} and by one more integration, we obtain
\begin{equation}\label{phi1}
\phi(\chi)=\frac{1}{\sqrt[5]{36}}\int{{F(\chi)}^{\frac 2 5}d \chi}\equiv Z~.
\end{equation}
Lastly, using the second equation of the system~\eqref{triad2}, together with  \eqref{lam2}, we find
\begin{equation}\label{mu1}
\mu(\chi)=\frac{6({F}_{\chi})^2-5 F{F}_{\chi\chi}}{50 {F}^3}~.
\end{equation}
Since we know $\lambda$ and $\mu$, the free  term of~\eqref{ansa3} becomes
\begin{equation}\label{las}
S(\chi)=\frac{6(F \mu ^2 - \mu_{\chi \chi})}{F\lambda^2}~.
\end{equation}
For the Kustaanheimo-Qvist function~\eqref{qvist}, we  find
\begin{equation}\label{SetC}
\begin{array}{l}
	\lambda=\sqrt[5]6 \sqrt{\alpha_1 x^2+\alpha_2 x+\alpha_3}~,\\
	\mu=-\frac 1 8 ({\alpha_2}^2-4 \alpha_1\alpha_3) \sqrt{\alpha_1 x^2+\alpha_2 x+\alpha_3}~,
\end{array}
\end{equation}
which leads to the constant $S$ given by
\begin{equation}\label{es10}
S=-\frac{\sqrt[5]{27}({\alpha_2}^2-4 \alpha_1\alpha_3)^2}{32 \sqrt[5]4}
\end{equation}
and implies that the general solution is a Weierstrass elliptic function.
The solutions of~\eqref{le} are  free from movable points only when the function $S(\chi)$ is of the linear form,
which by a trivial change of the variables takes one of the following three standard forms:
\begin{align}\label{triad5}
\left\{
\begin{array}{lll}
S(\chi)&=0 \Rightarrow W_{ZZ}=6 W^2~, \\
S(\chi)&=\frac 1 2  \Rightarrow W_{ZZ}=6 W^2 +\frac 12~,  \\
S(\chi)&=\phi(\chi)=Z \Rightarrow W_{ZZ}=6 W^2+Z.
\end{array}
\right.
\end{align}
If $S$ is not a constant, the general solution is a Painlev\'e transcendental function. The last case in (\ref{triad5}) is usually known as $P_I$, the first of the six classes of Painlev\'e type equations. For more details on this Painlev\'e pattern of solutions, with application to non-static radially symmetric distributions of matter in general relativity, we refer the reader to Wyman's Jeffery-Williams lecture, 1976 \cite{wyman}.

 %%%%%%%%%%%%%%%%%%%%%%%%%%%%%%%%%%%%%%
\subsection{The $n=2$ Emden-Fowler equation of the first class}
%%%%%%%%%%%%%%%%%%%%%%%%%%%%%%%%%%%%%%
Following our setting, we will study in detail the Emden-Fowler equation
\begin{equation}\label{nep}
\frac{d^2z}{d \chi^2}=\frac 3 2 \chi^{-\frac 52}z^2~.
\end{equation}
 The  integral of motion is obtained from~(\ref{100})  using $n=2$ and $A=\frac 3 2$ to give
  \begin{equation}\label{eqa1}
\frac{dz}{d\chi}\left(\chi \frac{dz}{d\chi}-z\right)-\left(\frac{z}{\sqrt \chi}\right)^3={\mathcal C_1}~,
 \end{equation}
which has a rational solution provided that ${\mathcal C_1}=0$, given by
\begin{eqnarray}\label{eqa2}
\begin{array}{ll}
z_p(\chi)=-\frac{K_1\chi}{(K_1+ \sqrt \chi)^2}~.
\end{array}
\end{eqnarray}
The particular solution of~(\ref{nep}), which is also given by~(\ref{er300tris}),  is exactly~(\ref{eqa2}).

To find the general solutions, we use the system~(\ref{equ2}) with  $n=2$, and we obtain
\begin{eqnarray}\label{equ2a2}
\begin{array}{ll}
&\chi(\tau)=a B_{1}^{\,2}\exp \Theta(\tau)\\
&z(\tau)=b B_1 \tau \exp{\frac{1}{2}\Theta (\tau)}\\
&\Theta(\tau)=\int{\frac{d \tau}{\sqrt{K_3+\frac{2\tau^3}{3}+\frac{\tau^2}{4}}}}~\\
\end{array}
\end{eqnarray}
where $\frac {a}{b^2}=\frac 9 4$. By redefining  $B_1=\frac{2B_2}{3 b}$, the parametric general solutions to~(\ref{nep}) become
\begin{eqnarray}\label{equ2a20}
\begin{array}{ll}
&\chi(\tau)={B_2}^2\exp {\Theta(\tau)}\\
&z(\tau)=\frac 2 3 B_2 \tau \exp {\frac{\Theta (\tau)}{2}}~.
\end{array}
\end{eqnarray}
 $\Theta(\tau)$ is obtained by inverting the elliptic equation
\begin{equation}\label{xo}
\left(\frac{d \tau}{d \Theta}\right)^2=\frac {2 \tau^3} {3}+\frac {\tau^2}{4}+K_3=a_3 \tau^3+a_2\tau^2+a_1 \tau+a_0\equiv Q_3(\tau)~,
\end{equation}
which is put in standard form using the scale-shift transformation
\begin{equation}\label{eq5}
\tau(\Theta)=\frac{4}{a_3}\wp(\Theta;g_2,g_3)-\frac{a_2}{3a_3}=6\wp(\Theta;g_2,g_3)-\frac 18
\end{equation}
to become
\begin{equation}\label{eq4}
\wp_{\Theta}^{\,\,2}=4 \wp^3-g_2 \wp -g _3~.
\end{equation}
The germs  of the Weierstrass function are given by
\begin{equation}\label{eq6}
\begin{array}{l}
g_2=\frac{a_2^2-3 a_1a_3}{12}=\frac{1}{192}\\
g_3=\frac{9 a_1a_2a_3-27 a_0 a_3^2-2 a_2^3}{432}=-\frac{1+384 K_3}{13824}
\end{array}
\end{equation}
and together with the modular discriminant
\begin{equation}\label{eq7}
\Delta=g_{2}^{\,3}-27 g_{3}^{\,2}=-\frac{K_3(1+192 K_3)}{9216}
\end{equation}
are used to classify the solutions of~(\ref{eq4}) \cite{Abra}.

\begin{description}
	\item [Case (1).]
If $ \Delta  \equiv 0 \Rightarrow K_3=0  $ or $K_3=-\frac{1}{192}$, the Weierstrass solutions can be simplified since $\wp$ degenerates into {\it hyperbolic} or {\it trigonometric} functions.
		\begin{enumerate}
		 \item [Case (1a).] $K_3=0\Rightarrow g_2>0,~g_3<0$, so~(\ref{eq4}) has soliton solution
\begin{equation} \label{eq13}
\tau(\Theta)=\frac 3 8 \left[ -1+ \mathrm{tanh}^2 \left(\frac {\Theta-\Theta_0}{ 4} \right)\right]~.
\end{equation}
Let
\begin{equation}\label{eq120}
\begin{array}{l}
	g_2=12{\hat e}^2>0\\
	g_3=-8{\hat e}^3<0~.\\
\end{array}
\end{equation}
 The Weierstrass $\wp$ solution to~(\ref{eq4}) reduces to
\begin{equation} \label{eq130}
\wp(\Theta;12{\hat e}^2,-8{\hat e}^3)=\hat e+3\hat e~ \mathrm{csch}^2(\sqrt{3\hat e}\Theta)~.
\end{equation}
Since  $\hat e=\frac{1}{48}>0$, the Weierstrass solution is
\begin{equation} \label{eq101}
\wp(\Theta)=\frac{1}{16}\left[\frac 1 3 +\mathrm{csch}^2 \left(\frac {\Theta-\Theta_0}{4} \right)\right]~,
\end{equation}
and using (\ref{eq5}) we obtain
\begin{equation} \label{eq102}
\tau(\Theta)=\frac 3 8 \mathrm{csch}^2 \left(\frac {\Theta-\Theta_0}{4} \right).
\end{equation}
		\item [Case (1b).] $K_3=-\frac{1}{192}\Rightarrow g_2>0,~g_3>0$, so~(\ref{eq4}) has the periodic solution
\begin{equation} \label{eq14}
\tau(\Theta)=\frac 1 8 \left[1+3 \mathrm{tan}^2 \left(\frac {\Theta-\Theta_0}{4} \right)\right]~.
\end{equation}

Let
\begin{equation}\label{eq140}
\begin{array}{l}
	g_2=12{\tilde {e}}^2>0~,\\
	g_3 =8{\tilde {e}}^3>0~.\\
\end{array}
\end{equation}
 The Weierstrass $\wp$ solution reduces to
\begin{equation} \label{eq150}
\wp(\Theta;12{\tilde {e}}^2,8{\tilde {e}}^3)=-{\tilde {e}}+3{\tilde {e}}~ \mathrm{csc}^2(\sqrt{3{\tilde {e}}}\Theta)~.
\end{equation}
Since  $\tilde e=\frac{1}{48}>0$, the Weierstrass solution gives
\begin{equation} \label{eq103}
\wp(\Theta)=\frac{1}{16}\left[-\frac 1 3 +\mathrm{csc}^2 \left(\frac \Theta 4 \right)\right]
\end{equation}
and using (\ref{eq5}) we obtain
\begin{equation} \label{eq104}
\tau(\Theta)=-\frac 1 4 \left[1-\frac 3 2 \mathrm{csc}^2 \left(\frac {\Theta-\Theta_0}{4} \right)\right]~.
\end{equation}
	\end{enumerate}
\item [Case (2).]
If $\Delta  \ne 0$, in general the Weierstrass solutions cannot be simplified, with the exception of a  particular lemniscatic case for which $g_2>0, g_3=0 \Rightarrow K_3=-\frac{1}{384}$.  In this case, we obtain
\begin{equation}\label{eq15}
\tau(\Theta)=-\frac 1 8+6\wp\left(\Theta-\Theta_0;\frac{1}{192},0\right)~.%=\frac 1 8\left[-1+2\sqrt 3\wp\left(\frac{\Theta}{\sqrt[4]{192}},1,0\right)\right]
\end{equation}
%%%%%
 Because $\Delta=\frac{1}{7077888}>0$, the cubic polynomial $4t^3-g_2 t$ has three distinct real roots given by $e_3=-\frac{\sqrt {g_2}}{2}$, $e_2=0$, and $e_1=\frac{\sqrt {g_2}}{2}$. Although the Weierstrass unbounded function  has poles aligned on the real axis of the $\Theta- \Theta_0$ complex plane, we can choose $\xi_0$ in such a way to shift these poles a half period above the real axis so that the elliptic function simplifies using the formula
\begin{equation}\label{sol5}
\wp(\Theta;g_2,0)=e_3+(e_2-e_3)\mathrm{sn}^2\left(\sqrt{e_1-e_3}(\Theta-\Theta_0);m\right)
\end{equation}
 with elliptic modulus $m=\sqrt{\frac{e_2-e_3}{e_1-e_3}}$.
Using the values of the roots, we obtain
\begin{equation}\label{sol6}
\wp(\Theta;g_2,0)=-\frac{\sqrt{g_2}}{2}\mathrm{cn}^2
\left(\sqrt[4]{g_2}\Theta;\frac{\sqrt{2}}{2}\right)~,
\end{equation}
which becomes
\begin{equation}\label{sol7}
\wp\left(\Theta;\frac {1}{192},0\right)=-\frac{1}{16 \sqrt 3}\mathrm{cn}^2\left(\frac{1}{2 \sqrt2 \sqrt[4]{3}}\Theta;\frac{\sqrt{2}}{2}\right)~.
\end{equation}
Using these results, the solution (\ref{eq15}) is
\begin{equation}\label{sol8}
\tau(\Theta)=-\frac{1}{8}\left[1+\sqrt{3}\mathrm{cn}^2\left(\frac{1}{2 \sqrt2 \sqrt[4]{3}}(\Theta-\Theta_0);\frac{\sqrt{2}}{2}\right) \right]~.
\end{equation}
\end{description}
For all  other values of $K_3$, one obtains the solutions in terms of the general $\wp$ functions, which take the form
\begin{equation}\label{eq50}
\tau(\Theta)=-\frac 1 8 +6\wp\left(\Theta-\Theta_0;\frac{1}{192},-\frac{1+384 K_3}{13824}\right)~.
\end{equation}

By inverting all of the above solutions in order (and keeping the only real and positive branches with positive argument), we obtain
%... 71
\begin{equation}\label{set1}
%\begin{array}{ll}
\Theta(\tau)=\left\{ \begin{array}{l}
4 ~\mathrm{arctanh}\left(\sqrt{1+\frac {8\tau}{3}}\right)+\Theta_0\\
4 ~\mathrm{arccsch}\left(2\sqrt{\frac {2\tau}{3}}\right)+\Theta_0\\
4 ~\mathrm{arctan}\left(\sqrt{\frac {8\tau-1}{3}}\right)+\Theta_0\\
4 ~\mathrm{arcsin}\left(\sqrt{\frac{3}{2(1+4 \tau)}}\right)+\Theta_0\\
2 \sqrt 2 \sqrt[4]3\,F\left(\mathrm{arccos}\left(\frac{\sqrt{-(8 \tau+1)}}{\sqrt[4]3}\right);\frac{\sqrt{2}}{2}\right)+\Theta_0\\
\wp^{-1}\left(\frac 1 6 \left(\frac 1 8 +\tau\right); \frac{1}{192},-\frac{1+384K_3}{13824}\right)+\Theta_0~,\\
\end{array} \right.
\end{equation}
where $F(\tau;m)$ is the elliptic integral of the first kind.

Using~\eqref{lam2}-\eqref{mu1}  we have
\begin{equation}\label{SetA}
\begin{array}{l}
	\lambda=\sqrt[5]4 \sqrt x~,\\
	\phi =\frac{1}{\sqrt[5]16}\ln x~,\\
	\mu=-\frac{\sqrt x}{2}~.
\end{array}
\end{equation}
By using \eqref{las}, this  case reduces to the elliptic equation
\begin{equation}
W_{ZZ}=6 W^2+\frac{1}{36\sqrt[5]{16}}~.
\end{equation}

%%%%%%%%%%%%%%%%%%%%
\subsection{The $n=2$ Emden-Fowler equations of the second class}
%%%%%%%%%%%%%%%%%%%%
Here, we consider our solution method for the Emden-Fowler equation
\begin{equation}\label{eqkq1}
\frac{d^2z}{d\chi^2}=-\chi^{-5}z^2~.
\end{equation}

Using the transformation  $z(\chi)=\frac{w(s)}{s}$, where $s=\frac 1 \chi$,  leads to
%..........................
\begin{equation}\label{eq001}
\frac{d^2w}{ds^2}=-w^2~,
\end{equation}
whose integral of motion is
\begin{equation}
\left(\frac{dw}{ds}\right)^2+\frac 2 3  w^3=\mathcal C_3~.
\end{equation}
In terms of the original variables, this invariant becomes
\begin{equation}
\chi^4 \left[\frac{d}{d \chi}\left(\frac{z}{\chi}\right)\right]^2+\frac 2 3 \left(\frac{z}{\chi}\right)^3=\mathcal C_3~,
\end{equation}
Using $A=-1$  with $n=2$, a particular solution ca be found using (\ref{er300tridd}), which gives
\begin{equation}\label{mo1}
z_p(\chi)=-\frac{6 x^3}{(1+K_2 x)^2}~.
\end{equation}
The same solution verifies the invariant when $\mathcal C_3=0$.

To find the general solutions, we use the  system~(\ref{equ5}),  which gives
\begin{eqnarray}\label{equ5a}
\begin{array}{ll}
&\chi(\tau)=\frac{a C_1}{\Theta(\tau)}\\
&z(\tau)=\frac{b C_{1}^{\,3}\tau}{\Theta(\tau)}\\
&\Theta(\tau)=\Theta_0+\int{\frac{d \tau}{\sqrt{1\pm \tau^ {3}}}}~.\\
\end{array}
\end{eqnarray}
Since $A=-1$, $a^3=\mp \frac 2 3 b$, and by redefining  $\pm aC_1=C_2$, the parametric general solutions of~(\ref{eqkq1}) are found from the system
\begin{eqnarray}\label{eqz5a}
\begin{array}{ll}
&\chi(\tau)=\frac{C_2}{\Theta(\tau)}\\
&z(\tau)= \frac {3{C_2}^3}{2}\frac{\tau}{\Theta(\tau)}~,
\end{array}
\end{eqnarray}
where  $\tau(\Theta)$ satisfies the reduced elliptic  equation
\begin{equation}\label{xo0}
\left(\frac{d \tau}{d \Theta}\right)^2=\pm  \tau^3+1~,
\end{equation}
which is a particular case $a_3=\pm 1$, $a_2=a_1=0$, and $a_0=1$ of~(\ref{xo}). The Weierstrass germs are $g_2=0$ and $g_3=-\frac {1}{16}$; therefore, $\Delta=-\frac{3^3}{2^8}
<0$ and the solution to~(\ref{xo0}) is given by the simplified equi-anharmonic case, and takes the form  % case, see Fig. \ref{figure3}.
\begin{equation}\label{xo1}
\tau(\Theta)=\pm 4 \wp \left(\Theta-\Theta_0; 0,-\frac{1}{16}\right)~.
\end{equation}
Inverting and using~(\ref{eqz5a}), we have
\begin{eqnarray}\label{eqz5b}
\begin{array}{ll}
&\chi(\tau)=\frac{C_2}{\Theta_0+\wp^{-1}\left(\pm \frac \tau 4; 0,-\frac{1}{16}\right)}\\
&z(\tau)= \frac {3{C_2}^3}{2}\cdot \frac{\tau}{\Theta_0+\wp^{-1} \left(\pm \frac \tau 4; 0,-\frac{1}{16}\right)}~.
\end{array}
\end{eqnarray}

Using~\eqref{lam2}-\eqref{mu1},  we have
\begin{equation}\label{SetB}
\begin{array}{l}
	\lambda=-\sqrt[5]{6} x~,\\
	\phi =-\frac{1}{\sqrt[5]36}\frac 1 x~,\\
	\mu=0~.
\end{array}
\end{equation}
By using \eqref{las}, this case reduces to the elliptic equation
%...103
\begin{equation}
W_{ZZ}=6 W^2~.
\end{equation}

%%%%%%%%%%%%%%%%%%%%%%%%%%%%%%%%%%%%%%%
 \subsection{The $n=5$ Emden-Fowler equation of the first class}
%%%%%%%%%%%%%%%%%%%%%%%%%%%%%%%%%%%%%%%
In 1907, Emden reviewed the theory of polytropic gas spheres (stars in astrophysics) and studied equations of the form \cite{Emden}
\begin{equation}\label{eq17a}
\frac{d^2z}{d\chi^2}=\pm \chi^{1-n}z^n~,
\end{equation}
which belongs to the first class if one chooses $n=5$. Letting  $A=1$ in~(\ref{ef2}), we obtain
\begin{equation}\label{eq16}
\frac{d^2z}{d\chi^2}=\chi^{-4}z^5~.
\end{equation}
This equation was first derived  by Schuster \cite{Sch}
and has the  integral of motion
\begin{equation}\label{eq141}
\frac{dz}{d\chi}\left(\chi \frac{dz}{d\chi}-z\right)-\frac 1 3 \left(\frac{z^2}{\chi}\right)^3=\mathcal{C}_1
\end{equation}
with particular solutions
\begin{equation}\label{eqa21}
\begin{array}{l}
z_p(\chi)=\pm \frac{\sqrt 6 K_5\chi}{\sqrt{{K_5}^4-12 \chi^2}}~\\
z_p(\chi)=\mp \frac{\sqrt 6 K_5\chi}{\sqrt{ \chi^2-12 {K_5}^4}}~,
\end{array}
\end{equation}
obtained when  $\mathcal{C}_1=0$.
The second  particular solution can also be obtained by using~(\ref{er300tris}) with $A=1,n=5$, and $K_1=12 {K_5}^4$.

To find the general solutions, we use the  system~(\ref{equ2}) with $n=2$ to obtain
\begin{eqnarray}\label{equ222}
\begin{array}{ll}
&\chi(\tau)=a B_{1}^{\,2}\exp \Theta(\tau)~,\\
&z(\tau)=b B_1 \tau \exp{\frac{1}{2}\Theta (\tau)}~,\\
&\Theta(\tau)=\int{\frac{d \tau}{\sqrt{K_3+\frac{\tau^6}{3}+\frac{\tau^2}{4}}}}~,
\end{array}
\end{eqnarray}
where $a=\pm b^2$. Redefining $bB_1=B_2$, the parametric general solutions of~(\ref{eq16}) become
%...109
\begin{eqnarray}\label{equ2222}
\begin{array}{ll}
&\chi(\tau)=\pm B_2 \exp \Theta(\tau)\\
&z(\tau)=B_2 \tau \exp{\frac{1}{2}\Theta (\tau)}~,
\end{array}
\end{eqnarray}
from which we can see that the solutions can be expressed  in terms of Weierstrass elliptic functions.
$\Theta(\tau)$ is obtained  by inverting  the elliptic equation
\begin{equation}\label{xo2}
\left(\frac{d \tau}{d \Theta}\right)^2=\frac{\tau^6}{3}+\frac{\tau^2}{4}+K_3~,
\end{equation}
which is
 \begin{equation}\label{eq301}
\left(\frac{d u}{d\Theta}\right)^2=u^4+u^2+K_4u=b_4u^4+b_3 u^3+b_2u^2+b_1u+b_0\equiv Q_4(u)~,
 \end{equation}
  using ${\tau}^2=\frac {\sqrt{3}}{2}u$ and $K_4=\frac{8 \sqrt 3}{3}K_3$. According to Whittaker and Watson \cite{Whi}, the solutions of~(\ref{eq301}) can be expressed by means of the rational Weierstrass elliptic functions $\wp(\Theta;g_2,g_3)$ using the formula
%\begin{equation}\label{eq5a}
\begin{widetext}
\begin{equation}\label{eq5a}
u(\Theta)=u_0+
%\begin{equation}\label{eq5a}
\frac{ \sqrt{Q_4(u_0)}\wp\rq{}(\Theta;g_2,g_3)+\frac 1 2 \frac{dQ_4}{d u}(u_0)\Big(\wp(\Theta;g_2,g_3)-\frac {1}{24}\frac{d^2 Q_4}{du^2}(u_0)\Big)+\frac{Q_4(u_0)}{24}%Q_4(u_0)
\frac{d^3Q_4}{d u^3}(u_0)}{2\Big(\wp(\Theta;g_2,g_3)-\frac{1}{24}\frac{d^2Q_4}{du^2}(u_0)\Big)^2-\frac{Q_4(u_0)}{48}%Q_4(u_0)
\frac{d^4Q_4}{du^4}(u_0)}~,
\end{equation}
\end{widetext}
where $u_0$ can be any constant and not necessarily a root of $Q_4(u)$. The  elliptic invariants and modular discriminant  are related to the coefficients of the quartic polynomial $Q_4(u)$, which yield
\begin{equation}\label{es3}
\begin{array}{l}
g_2=\frac{1}{12}\left({b_2}^2-3 b_1b_3\right)+b_0b_4=\frac{1}{12}\\
g_3=\frac{1}{432}\left[9b_2(b_1b_3+8 b_0b_4)-2 {b_2}^3-27\left(b_0{b_3}^2+{b_1}^2b_4\right)\right]\\
\,\quad=-\frac{2+27 {K_4}^2}{432}\\
\Delta={g_{2}}^3-27 {g_{3}}^2=-\frac{{K_4}^2(4+27 {K_4^2})}{256}~,
\end{array}
\end{equation}
and  are used to classify  the Weierstrass solutions.
Because we already know a simple root $u_0$ of $Q_4(u)$,~(\ref{eq5}) has the simpler form
 \begin{equation}\label{eq6a}
u(\Theta)=u_0+\frac{\frac{d Q_4}{du}(u_0)}{4 \Big(\wp(\Theta;g_2,g_3)-\frac {1}{24}\frac{d^2 Q_4}{d u^2}(u_0)\Big)}~.
\end{equation}
 Since $\Delta<0$, $g_2>0$, and $g_3<0$, the Weierstrass solutions cannot be simplified further, and for $u_0=0$ we have the solution
 \begin{equation}\label{eq7a}
u(\Theta)=\frac{K_4}{4\wp\left(\Theta-\Theta_0;\frac{1}{12},-\frac{2+27 {K_4}^2}{432}\right)-\frac {1}{3}}~, \qquad K_4\ne0~.
\end{equation}
Using back the transformations, we obtain the solution
\begin{equation}\label{eq8b}
\Theta(\tau)=\Theta_0+\wp^{-1}\left(\frac{K_3}{\tau^2}+\frac{1}{12}; \frac {1}{12}, -\frac{1+288 {K_3}^2}{216}\right), \quad K_3\ne 0~.
\end{equation}
When $K_4=0$,~(\ref{eq301}) becomes
\begin{equation}\label{eq9b}
\frac{du}{u\sqrt{u^2+1}}=\pm d \Theta~,
\end{equation}
which gives
\begin{equation}\label{eq10b}
\Theta(\tau)=\Theta_0\pm \ln\left({\frac{2\tau^2}{\sqrt 3+\sqrt{3+4 \tau^4}}}\right)~, \,\, K_4=0~.
\end{equation}
Notice that this particular solution leads to the solutions given by system~(\ref{eqa21}) when $\Theta_0=0$.
Solutions in terms of Jacobian and Weierstrass elliptic functions for this important case in astrophysics have been obtained only recently by Mach~\cite{pmach}.

\bigskip
\section{Conclusion}
In summary, we have analyzed the two integrable classes of Emden-Fowler equations for which the power parameters of the independent and dependent variables are related through $\lambda=(n-1)/2$ and $\lambda=n+1$. For these cases, the parametric solutions were written explicitly following Polyanin and Zaitsev \cite{Pol}.
As particular examples, we have presented  the astrophysical case $\lambda=2$, $n=5$ belonging to the first class, and two $n=2$ cases that describe perfect fluids in general relativity, one with $\lambda=1/2$ belonging to the first class and another one with $\lambda=3$ from the second class, including their solutions in terms of  Weierstrass elliptic functions or simpler reductions thereof. The $n=2$ Emden-Fowler equations have been also presented as particular cases of an equation to which Ince's method of Painlev\'e reduction can be applied.

Since   in general we  cannot obtain closed form  solutions from  parametric  solutions (some parameters  are impossible to eliminate), we shall  resort to calculating  invariants or using transformations that represent an analog of classical invariant theory. This  will make the new  equations integrable without using the two conditions between $\lambda$ and $n$, since  the case $n=2$ can be always reduced to one of three  Painlev\'e transcendents. Finally, we noticed that there are many other cases with potential applications, such as those with negative $n$ such as Ermakov equation, that  belong to these two integrable classes, and that can be approached along the lines presented in this paper.

 \bigskip
 \bigskip

{\bf Acknowledgements}

\medskip

S.C. Mancas would like to acknowledge partial support from the Dean of Research \&
Graduate Studies at ERAU while on a short visit to IPICyT,  San Luis Potos\'{\i}, Mexico.
%The authors wish to thank the anonymous referee for pointing them the connection with the Painlev\'e equations.

{\small
%\section*{References}

}


\begin{thebibliography}{10}
%1
\bibitem{Emden} R. Emden,
\textit{Gas Balls: Applications of the Mechanical Heat Theory to Cosmological and Meteorological Problems} (Teubner, Berlin, 1907) [in German].
%2
\bibitem{Lane} J.H. Lane,
``On the theoretical temperature of the sun under the hypothesis of a gaseous mass maintaining its volume by its internal
heat and depending on the laws of gases known to terrestrial experiment",
Am. J. Sci. {\bf 50}, 57-74 (1870).
%3
\bibitem{Edd} Sir A.S. Eddington,
\textit{The Internal Constitution of the Stars} (Cambridge Univ. Press, Cambridge, 1926).
%4
\bibitem{Fow30} R.H. Fowler,
``The solutions of Emden\rq{}s and similar differential equations",
 Mon. Not. R. Astron. Soc. {\bf 91}, 63-91 (1930).
%5
\bibitem{Mil30} E.A. Milne,
``The analysis of stellar structure",
 Mon. Not. R. Astron. Soc. {\bf 91}, 4-52 (1930).
%6
\bibitem{RM54} A.W. Rodgers and D.M. Myers,
``Solutions of the negative Emden polytropes",
 Mon. Not. R. Astron. Soc. {\bf 114},  620-627 (1954).
%7
\bibitem{JM94} J. Janus and J. Myjak,
``A generalized EF equation with a negative exponent",
Nonlinear Anal. Theory. Methods. Appl. {\bf 23}, 953-970 (1994).
%8
\bibitem{B53} R. Bellman,
\textit{Stability Theory of Differential Equations}, pp. 143-144 (McGraw-Hill, New York, 1953).
%9
\bibitem{Thomas} L.H. Thomas,
``The calculation of atomic fields",
Math. Proc. Camb. Philos. Soc. {\bf 23}, 542-548 (1927).
%10
\bibitem{Fermi} E. Fermi,
``A statistical method for the determination of some atomic properties",
Rend. Accad. Naz. Lincei {\bf 6}, 602-607 (1927) [in Italian].
%11
\bibitem{gh} H. Goenner and P. Havas,
``Exact solutions of the generalized Lane-Emden equation",
J. Math. Phys. {\bf 41}, 7029-7042 (2000).
%12
\bibitem{hm} T. Harko and M.K. Mak,
``Exact power series solutions of the structure equations of the general relativistic
isotropic fluid stars with linear barotropic and polytropic equations of state",
Astrophys. Space Sci. {\bf 361}: 283, 1-19 (2016).
%13
\bibitem{Bhut} O.P. Bhutani and K.  Vijayakumar, ``On certain new and exact solutions of the Emden-Fowler equation and Emden equation via invariant variational principles and group invariance'', J. Aust. Math. Soc. Ser. B. {\bf 32}(4), 457-468 (1991).
%14
\bibitem{Pana1} D.E. Panayotounakos, and N.  Sotiropoulos, ``Exact analytic solutions of unsolvable classes of first-and second-order nonlinear ODEs (Part II: Emden-Fowler and relative equations)'', Appl. Math. Lett. {\bf 18}(4), 367-374  (2005).
%15
\bibitem{Pana2} D.E. Panayotounakos, and D.C.  Kravvaritis, `` Exact analytic solutions of the Abel, Emden-Fowler and generalized Emden-Fowler nonlinear ODEs'', Nonlinear Anal. Real World Appl. {\bf 7}(4), 634-650 (2006).
%16
\bibitem{Car} J.F. Cari{\~n}ena,  P.G.L. Leach, and J.  De Lucas, ``Quasi-Lie schemes and Emden-Fowler equations'', J. Math. Phys. {\bf 50}(10), 103515 (2009).
%17
    \bibitem{Blu} G. Bluman and S. Anco,
\textit{Symmetry and integration methods for differential equations} (Springer Science \& Business Media, Berlin, 2008, Vol. 154).
%18
\bibitem{Pete} P.J. Olver,
 \textit{Applications of Lie Groups to Differential Equations} (Graduate Texts in Mathematics, No. 107. Springer-Verlag, New York, 1986).
%19
\bibitem{LMM92} P.G.L. Leach, R. Maartens, and S.D.  Maharaj, ``Self-similar solutions of the generalized Emden-Fowler equation'', Int. J. Non-Linear Mech. {\bf 27}(4), 575-582 (1992).
%20
\bibitem{Fow14} R.H. Fowler,
``The form near infinity of real continuous solutions of a certain differential equation of the second order",
Quart. J. Math. (Oxford) {\bf 45}, 289-350 (1914).

%\bibitem {Kamke} E. Kamke,
%\textit{Differentialgleichungen. L\"osungsmethoden und L\"osungen I. Gew\"ohnliche Differentialgleichungen},
%eqs. 6.11 at p. 544 and  6.74 at p. 560 (Chelsea Publ. Company, New York, 1948).
%21
\bibitem{ros} P. Rosenau,
``A note on the integration of the Emden-Fowler equation",
Int. J. Non-Linear Mech. {\bf 19}, 303-308 (1984).
%22
\bibitem{Djuk} D.S. Djukic,
``A procedure for finding first integrals of mechanical systems with gauge-variant Lagrangians",
Int. J. Non-Linear Mech. {\bf 8}, 479-488 (1973).
%23
\bibitem{Fow} R.H. Fowler,
``Further studies of Emden's and similar differential equations",
Quart. J. Math. (Oxford) {\bf os-2}(1), 259-288 (1931).
%24
\bibitem{ro} H.C. Rosu and O. Cornejo-P\'erez,
``Supersymmetric pairing of kinks for polynomial nonlinearities",
Phys. Rev. E {\bf 71}, 046607 (2005).
%25
\bibitem{Pol} A.D. Polyanin and V.F. Zaitsev,
\textit{Handbook of exact solutions for ordinary differential equations} (CRC Press Company, Boca Raton, 1995).
%26
\bibitem{KQ} P. Kustaanheimo and B. Qvist,
``A note on some general solutions of the Einstein field equations in a spherically symmetric world",
Societas Sci. Fennica. Comm. Physico-Mathematicae {\bf XIII.16}, 1 (1948). Reprinted in Gen. Rel. Grav. {\bf 30}, 663-673 (1998).
%27
\bibitem{Ince} E.L. Ince,
\textit{Ordinary Differential Equations} (Dover Publications, Inc., New York, 1956) pp. 328-330.
%28
\bibitem{Pain} P. Painlev\'e,
``M\'emoire sur les \'equations diff\'erentielles dont l'int\'egrale g\'enerale est uniforme",
Bull. S.M.F. {\bf 28}, 201-261 (1900).
%29
\bibitem{wyman} M. Wyman,
``Jeffery-Williams lecture, 1976, Non-static radially symmetric distributions of matter",
Can. Math. Bull. {\bf 19}, 343-357 (1976).
%30
\bibitem{Abra} M. Abramowitz and I. Stegun,
\textit{Handbook of Mathematical Functions: with Formulas, Graphs, and Mathematical Tables} (Dover, New York, 1972).
%31
\bibitem{Sch} A. Schuster,
``On the internal constitution of the Sun",
Rep. Br. Ass. Advmt Sci., 427-429 (1883).
%32
\bibitem{Whi} E.T. Whittaker and G.N. Watson,
\textit{A Course of Modern Analysis} (Cambridge University Press, Cambridge, 1927).
%33
\bibitem{pmach} P. Mach,
``All solutions of the $n = 5$ Lane-Emden equation",  J. Math. Phys. {\bf 53}, 062503 (2012).

\end{thebibliography}
\end{document}